\documentclass[journal,two side,web]{ieeecolor}
\usepackage{amsmath,amsfonts}
\usepackage{algorithmicx}  
\usepackage[ruled,linesnumbered]{algorithm2e}
\usepackage{generic}
\usepackage{cite}
\usepackage{amsmath,amssymb,amsfonts}
\usepackage{graphicx}
\usepackage{textcomp}
\usepackage{bm}
\usepackage{threeparttable}
\usepackage{bbm}
\def\BibTeX{{\rm B\kern-.05em{\sc i\kern-.025em b}\kern-.08em
    T\kern-.1667em\lower.7ex\hbox{E}\kern-.125emX}}
\begin{document}
\title{Self-Calibrated Dual Contrasting for Annotation-Efficient Bacteria Raman \\ Spectroscopy Clustering and Classification}
\author{Haiming Yao, Wei Luo, Tao Zhou, Ang Gao, and Xue Wang, \IEEEmembership{Senior Member, IEEE}
\thanks{Manuscript received XX XX, 20XX; revised XX XX, 20XX. This study was supported in part by the Ministry of Industry and Information Technology of the People's Republic of China. (Grant No. 2023ZY01028). (Corresponding author: Xue Wang.)}
\thanks{The authors are with the State Key Laboratory of Precision Measurement Technology and Instruments, Department of Precision Instrument, Tsinghua University, Beijing 100084, China. (e-mails: $\{$yhm22, luow23, zhoutao24, ga24$\}$@mails.tsinghua.edu.cn; 
wangxue@mail.tsinghua.edu.cn).}}

\maketitle

\begin{abstract}
Raman scattering is based on molecular vibration spectroscopy and provides a powerful technology for  pathogenic bacteria diagnosis using the unique molecular fingerprint information of a substance. The integration of deep learning technology has significantly improved the efficiency and accuracy of intelligent Raman spectroscopy (RS) recognition. However, the current RS recognition methods based on deep neural networks still require the annotation of a large amount of spectral data, which is labor-intensive. This paper presents a novel annotation-efficient Self-Calibrated Dual  Contrasting (SCDC) method for RS recognition that operates effectively with few or no annotation. Our core motivation is to represent the spectrum from two different perspectives in two distinct subspaces: embedding and category. The embedding perspective captures instance-level information, while the category perspective reflects category-level information. Accordingly, we have implemented a dual contrastive learning approach from two perspectives to obtain discriminative representations, which are applicable for Raman spectroscopy recognition under both unsupervised and semi-supervised learning conditions. Furthermore, a self-calibration mechanism is proposed to enhance robustness. Validation of the identification task on three large-scale bacterial Raman spectroscopy datasets demonstrates that our SCDC method achieves robust recognition performance with very few (5$\%$ or 10$\%$) or no annotations, highlighting the potential of the proposed method for biospectral identification in annotation-efficient clinical scenarios.

\end{abstract}


\section{Introduction}
\label{sec:introduction}
\IEEEPARstart{R}{ecently}, the identification of pathogenic bacterial species and their sensitivity to antibiotics has garnered significant attention to enable accurate diagnosis of diseases caused by bacterial infections and to inform treatment decisions\cite{r38, r1}. Raman scattering, a potent qualitative and quantitative analysis technique based on molecular vibrations\cite{r15}, offers several advantages in the identification of pathogenic bacteria\cite{r12}. Notably, it bypasses the lengthy culture processes associated with traditional bacterial antibiotic sensitivity testing\cite{r3}, thereby facilitating early disease detection\cite{r1}. Moreover, this technique is non-destructive and non-invasive, eliminating the need for dyes or contrast agents\cite{r39}, and is applicable across diverse scientific fields, including chemistry\cite{r41} and medicine\cite{r40}. However, in practical applications, the complex matrix of the sample significantly influences the final spectrum\cite{r42}, complicating the accurate identification of the target substance's spectral information through subjective visual assessment. This challenge has necessitated the integration of analytical metrology to enhance the accuracy of spectral analysis.

More recently, data-driven learning-based approaches have garnered significant attention among various analytical tools, including Raman spectroscopy. For diagnosis-related recognition tasks, These methods can broadly be divided into two main categories: machine learning techniques, such as support vector machines (SVMs)\cite{r13} and random forests\cite{r43}, and deep learning methods that leverage deep neural networks, including convolutional neural networks (CNNs)\cite{r44,r34,r11} and long short-term memory (LSTM)\cite{r38}. While the qualitative and quantitative predictive models derived from these methods can achieve significantly higher accuracy than traditional analytical techniques and even surpass human performance, these learning methods typically employ a fully supervised approach, necessitating a substantial amount of well-annotated data to achieve satisfactory performance. Meanwhile, the manual annotation of Raman spectra is labor-intensive, particularly when dealing with large-scale datasets\cite{r45}. Consequently, it is imperative to develop annotation-efficient diagnosis methods that minimize the reliance on extensively annotated training datasets.

To overcome the above limitations of fully supervised methods, more promising solutions lie in semi-supervised classification or unsupervised clustering approaches, which require little or no annotations. Unsupervised clustering techniques automatically categorize spectra based on shared features by assessing the intrinsic properties of the spectra themselves without any annotation\cite{r45}. These methods analyze similarities and differences within the data and have been employed in existing studies for the diagnostic analysis of Raman spectra, including techniques such as $k$-means clustering\cite{r46}, hierarchical clustering\cite{r47}, and deep clustering\cite{r48}. Semi-supervised classification leverages partially labeled data alongside unlabeled data to enhance diagnostic performance, primarily relying on observed assumptions\cite{r49} to fully exploit the relationships within the data. While this approach has been extensively researched in the domains of natural or medical image analysis\cite{r50}\cite{r51}, it remains under-explored in the context of diagnostic analysis of Raman spectroscopy. Furthermore, the inherent low efficiency of Raman scattering often results in bacteria Raman spectra exhibiting poor signal-to-noise ratios(SNR)\cite{r12}, where subtle spectral differences can be obscured by noise. Additionally, high-dimensional bacteria Raman spectral datasets, characterized by a diverse range of categories, present complex data distribution manifolds\cite{r1}. These complexities pose significant challenges for Raman analysis when employing unsupervised clustering and semi-supervised classification methods with no or limited annotations.

To address the aforementioned challenges, we propose an annotation-efficient recognition method for bacterial Raman spectroscopy in this study: Self-Calibrated Dual Contrasting(SCDC). Specifically, SCDC initially constructs contrastive pairs through spectral augmentation. 
Then, it performs dual contrastive learning at both instance- and cluster-level for the feature representation matrix, with the goal of pulling closer positive pairs while pushing apart negative pairs, thereby enabling SCDC to effectively learn discriminative features. Furthermore, a self-calibration mechanism is proposed for the above-mentioned contrasting to further enhance robustness. Notably, the proposed method is capable of operating in both unsupervised and semi-supervised contexts and achieves state-of-the-art (SOTA) performance. We conducted experimental validation on three large-scale bacterial Raman spectroscopy datasets. On the well-known large-scale bacterial identification dataset \cite{r1}, our method achieved recognition accuracies of 70.8$\%$ and 73.8$\%$ using only 5$\%$ and 10$\%$ of the annotated data, respectively, outperforming the existing state-of-the-art methods by +10.2$\%$ and +2.5$\%$. On the bacterial strain \cite{r12} and marine pathogen bacteria \cite{r38} datasets, our method achieved recognition accuracies exceeding 90$\%$ using only 10$\%$ of the annotated data. Extensive experimental results demonstrate that SCDC significantly improves the intelligent Raman spectroscopy recognition performance with minimal annotation requirements. This approach demonstrates significant application potential in reducing the burden of expert annotation and contributing to more efficient diagnosis of diseases caused by bacterial infections.

\section{Related Works}
\subsection{Raman spectroscopy intelligent diagnosis}

The development of intelligent methods based on learning (artificial intelligence) has significantly enhanced the accuracy of automatic diagnosis in Raman spectroscopy. The researchers initially explored machine learning methods, \emph{e.g.}, discriminant analysis (DA) \cite{r52}, support vector machines (SVM)\cite{r53}, \emph{etc}. This approach typically employs principal component analysis (PCA) or other dimensionality reduction techniques, followed by the application of intelligent classification algorithms for diagnosis. In \cite{r55}, the authors proposed combining PCA with SVM to perform isocitrate dehydrogenase genotyping on glioma biopsy samples. In \cite{r54}, Raman spectroscopy analysis utilizing PCA and linear discriminant analysis (LDA) was implemented for the detection of coronary heart disease. Subsequently, with the advancement of deep neural networks, researchers began to focus on more robust deep learning methods. A notable milestone was the application of deep residual networks (ResNet) for bacterial classification using Raman spectroscopy in \cite{r1}. Following this, deep networks specifically tailored for Raman spectroscopy were proposed, including the scale-adaptive network (SANet)\cite{r16} and the RamanNet \cite{r56}. However, nearly all existing methods, including those mentioned above, depend on favorable data-driven conditions, specifically conventional supervised learning scenarios that require substantial amounts of spectral data and corresponding annotations.

\subsection{Unsupervised data clustering}

Unsupervised cluster analysis aims to identify high-density data clusters without any human annotation, ensuring that the data points within each cluster exhibit greater similarity while maintaining reduced similarities between different clusters. Due to their inherent advantage of not requiring annotations, numerous clustering-based unsupervised recognition methods have been developed, including hierarchical clustering\cite{r47}, $k$-means clustering\cite{r46}, as well as recent deep clustering techniques such as variational autoencoders\cite{r44} and deep embedding clustering\cite{r57}. These methods have been applied in Raman spectroscopy analysis; for example, \cite{r58} utilized PCA and hierarchical clustering to achieve Raman spectroscopy recognition of three waterborne pathogens, while the RamanCluster framework proposed in \cite{r45} was employed for unsupervised bacterial recognition.

\subsection{Semi-supervised classification}

\begin{figure*}[t]
\centerline{\includegraphics[width=150mm]{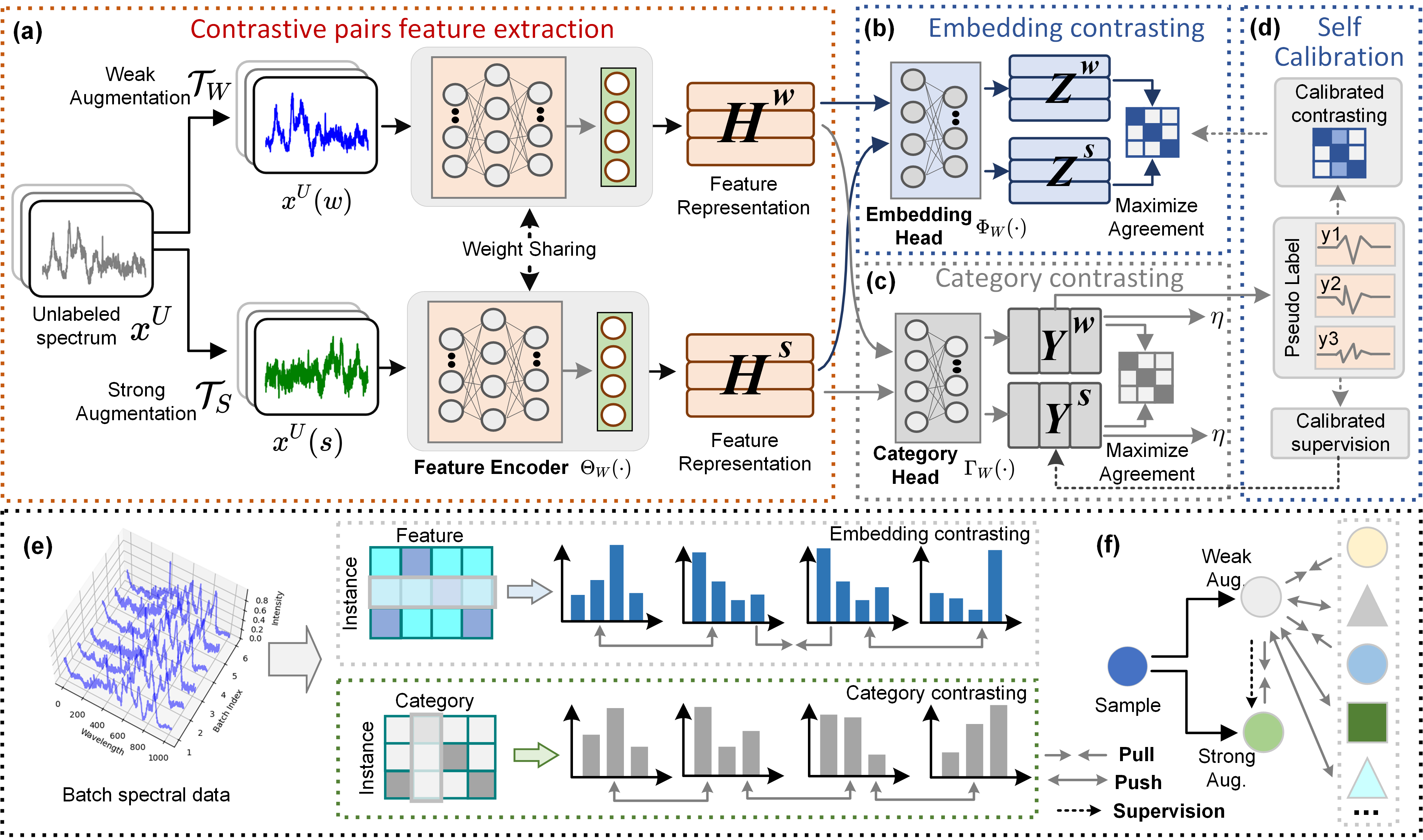}}
\caption[width=150mm]{Detailed framework of the proposed SCDC. (a). Construction of contrastive spectral pairs and extraction of their feature extraction process. (b). Embedding contrasting diagram, the feature representations are projected by the embedding head, then the row vectors are used for embedding contrastive learning. (c). Category contrasting diagram, the feature representations are projected by the category head, then the column vectors are used for category contrast learning. (d) The prediction results of weak augmentation are used as pseudo-labels to feedback into the aforementioned two contrasting processes for self-calibration. (e) Schematic diagram of the dual contrasting process for a batch of spectral data. (f) The self-calibration mechanism, where the prediction results of weakly augmented spectra are used to implement supervised contrastive learning and are also utilized for the pseudo supervision of strong augmented spectra. Blocks of different shapes represent spectra of different categories.}
\label{fig1}
\end{figure*}

Semi-supervised learning is an annotation efficient approach that requires only a portion of the data to be annotated. The functioning of this method is based on three core assumptions \cite{r49}: 1) the smoothness assumption, 2) the low-density assumption, and 3) the manifold assumption. This approach primarily encompasses two categories of methods\cite{r51}: inductive methods and transductive methods. It has been applied in various fields, including medical image classification\cite{r50} and fault diagnosis\cite{r59} in industrial settings. However,to the best of our knowledge, it has not yet been fully applied in the field of biological spectroscopy, including Raman spectroscopy.

\subsection{Contrastive representation learning}

Contrastive learning is a type of self-supervised learning method \cite{r60} that learns data representations by comparing pairs of samples. The core principle of contrastive learning is that similar samples should be close to one another in the representation space, while dissimilar samples should be farther apart. Notable methods in this domain include MoCo\cite{r61}, SimCLR\cite{r62}, BYOL\cite{r63}, and SimSiam\cite{r64}. This approach is widely utilized in the field of computer vision and has demonstrated the potential to surpass traditional supervised learning. In \cite{r65}, this approach was applied for transferable pre-training in Raman spectroscopy-based skin cancer diagnosis.

\section{Methodology}
\subsection{Preliminary:tasks and notations}

We begin by defining our task. In the intelligent recognition of Raman spectra, there are typically two datasets: a training set $\mathbf{D} _{train}$ and a test set $\mathbf{D} _{test}$. These datasets consist of $M+N$ pairs of samples and their corresponding annotations$\left \{ \left ( x_i,y_i  \right )  \right \}_{i=1}^{M+N}$, where 
$x_i\in \mathbb{R}^L$ represents a spectrum of length $L$, and $y_i\in \mathbb{R}$ denotes the category of the spectrum. The intelligent model is trained on the training set and evaluated on the test set. Generally, the acquisition of $y_i$ requires human annotation. To facilitate efficient annotation training, we annotate only a portion of the data in the training set, denoted as $\mathbf{D} _{train}^A=\left \{ \left ( x_i^A,y_i^A  \right )  \right \}_{i=1}^{M_A}$. The remaining data, which is not annotated, is denoted as $\mathbf{D} _{train}^U=\left \{ \left ( x_i^U  \right )  \right \}_{i=1}^{M_U}$. Therefore, in our setting, dataset $\mathbf{D}_{train}^A\cup \mathbf{D} _{train}^U$ with a reduced annotation burden, was utilized to finish the model training.

\subsection{Supervised Learning with annotated data}

The framework of the proposed method is illustrated in Fig. 1. It comprises three modules: a feature encoder $\Theta_{W1}\left ( \cdot  \right )$, an embedding head $\Phi_{W2}\left ( \cdot  \right ) $, and a category head $\Gamma _{W3}\left ( \cdot  \right )$. For labeled data, we utilize the encoder $\Theta_{W1}\left ( \cdot  \right )$ to obtain the representation of $x_i^A$, which is then passed to the category head $\Gamma _{W3}\left ( \cdot  \right )$ for category prediction:
 \begin{equation}
\tilde{y} _i^A = \Gamma _{W3}\left ( \Theta_{W1}\left ( x_i^A  \right )  \right ) 
\end{equation}
we employ cross-entropy loss for supervised training:
 \begin{equation}
\mathcal{L}_{sup}= \frac{1}{B} \sum_{i=1}^{B} \mathbf{H}\left (  y_i^A\mid  \tilde{y} _i^A\right )
\end{equation}
where the $\mathbf{H}\left (   \cdot \mid   \cdot \right )$ denotes the cross-entropy loss, and $B$ is the batch size.

\subsection{Unsupervised Learning with unannotated data}

We employ contrastive learning to leverage unlabeled training data $\mathbf{D} _{train}^U$, which primarily involves three processes: embedding contrasting, category contrasting, and self-calibration.

\subsubsection{Spectral augmentation for contrastive pairs}

Data augmentation is a fundamental component of contrastive learning methods, aiming to maximize the similarity between different views of the same sample while minimizing similarity with other samples. By generating views through different augmentations, we can enhance the robustness of the learned representations \cite{r66}. Therefore, we propose to use two types of augmentations: a weak spectral augmentation $\mathcal{T}_W$, and a strong spectral augmentation $\mathcal{T}_S$.

When enhancing Raman spectra, it is crucial to consider their physical significance, as the vibration intensity at each wavelength reflects the substance’s information. Therefore, for weak spectral augmentation $\mathcal{T}_W$, we introduce small local changes to the signal without altering the overall characteristics of the spectrum or significantly modifying its shape. This augmentation involves local perturbations, such as adding random Gaussian noise and applying random scale transformations \cite{r16}. In contrast, strong augmentation $\mathcal{T}_S$ manipulates the global context of the spectrum, further introducing spectral shifts and spectral deflections on the basis of the weak augmentation \cite{r65}.

Formally, for each input sample $x_i^U$ , we denote its strongly augmented view as $x_i^U(s)$ and its weakly augmented view as $x_i^U(w)$, where are sampled from two different augmentation families, i.e., $x_i^U(s)\sim \mathcal{T}_S$ and $x_i^U(w)\sim \mathcal{T}_w$. Subsequently, these views are then passed through the encoder $\Theta_{W1}\left ( \cdot  \right )$ to extract their high-dimensional latent representations as $h_i^s=\Theta_{W1}\left ( x_i^U(s)  \right ) $ and $h_i^w=\Theta_{W1}\left ( x_i^U(w)  \right )$. We record the corresponding representation matrix for a batch spectral data as $H^s$ and $H^w$ as shown in Fig.1 (a).

\subsubsection{Embedding contrasting}

First, we apply contrastive learning to each instance in order to explore its intrinsic properties. Specifically, we treat augmented versions of the same instance as positive pairs and other samples as negative pairs, aiming to maximize the similarity of positive pairs while minimizing the similarity of negative pairs. As shown in Fig.1 (b), to prevent potential information loss\cite{r62}, we employ the embedding head $\Phi_{W2}\left ( \cdot  \right ) $ to map the latent representation into the embedding space as $z_i^\Lambda=\Phi_{W2}\left ( h_i^\Lambda \right ),\Lambda=\left \{ s,w \right \}$, where contrastive learning is then performed. 

For a mini-batch of spectral data of size $B$, each instance $x_i^U$ is augmented and embedded using the aforementioned pipeline, resulting in $2B$ embeddings. We denote the embedding matrix of two batches as $Z^s$ and $Z^w$, for each embedding $z_i^\Lambda$, the embedding corresponding to the other augmented view of the same input is treated as the positive sample, forming a positive pair. Meanwhile, the remaining 
$2B-2$ embeddings from other inputs within the batch are considered negative samples, forming negative pairs. To this end,  we adopt the NT-Xent loss \cite{r62} to optimize the pair-wise similarities:
 \begin{equation}
\ell \left (z_i^s,z_i^w \right )  = -\mathrm{log} \frac{\mathrm{exp}\left ( \mathrm{sim} \left ( z_i^s,z_i^w \right ) /\tau \right )  }{\sum_{j=1}^{B}\mathbbm{1}_{[j\ne i]} \mathrm{exp}\left ( \mathrm{sim} \left ( z_i^s,z_j^\Lambda \right ) /\tau \right )  }    
\end{equation}
 \begin{equation}
\mathcal{L}_{emb} = \frac{1}{B} \sum_{i=1}^{B}\left [ \ell \left (z_i^s ,z_i^w\right )+  \ell \left (z_i^w,z_i^s \right )\right ] 
\end{equation}
where $sim(u,v)=uv^T/\left \| u \right \| \left \| v \right \| $ denotes the cosine similarity for row vector, $\mathbbm{1}_{[j\ne i]}\in \left \{ 0,1 \right \}$  is the indicator function, which equals 1 when  $j\ne i$, and $\tau$ is the temperature hyper-parameter that controls the softness.

\subsubsection{Category contrasting}

When the category head $\Gamma _{W3}\left ( h_i^\Lambda  \right )$ projects the representation $h_i^\Lambda $ into the subspace of sample category probabilities (after $\mathrm{soft}$-$\mathrm{max}$ activation function), the $k$-th element of the feature vector represents the probability that the sample belongs to the $k$-th category. Consider a mini-batch of data, as illustrated in Fig. 1 (c). From the column-wise perspective, each column vector represents the distribution of a category. Therefore, under different augmentations, the distribution of the same category should remain similar, while the distributions of different categories should differ. Based on this observation, we introduce the concept of category contrasting.

Specifically, when the batch size is $B$ and the number of categories is $C$, the probability matrices obtained from the two augmented batches are 
denotes as the $Y^s$ and $Y^w$, both of which have a shape of $\mathbbm{R}^{B\times C}$. We use $\bar{y} _i \in \mathbbm{R}^{B}$ to denote the $i$-th column vector of $Y$. For an column vector $\bar{y} _i$, the column vector corresponding to the other augmented view of the same category is treated as a positive sample, forming a positive pair, while the remaining $2C-2$ sample pairs are considered negative pairs. Similarly, the NT-Xent loss is used to perform category contrasting:
 \begin{equation}
 \bar{\ell } \left ( \bar{y}_i ^s,\bar{y} _i ^w \right ) =-\mathrm{log} \frac{\mathrm{exp}\left ( \mathrm{sim} \left ( \bar{y}_i ^s,\bar{y} _i ^w \right ) /\tau \right )  }{\sum_{j=1}^{C}\mathbbm{1}_{[j\ne i]} \mathrm{exp}\left ( \mathrm{sim} \left ( \bar{y} _i ^s,\bar{y} _j ^\Lambda \right ) /\tau \right )  }
\end{equation}
 \begin{equation}
\mathcal{L}_{cat} =\frac{1}{C} \sum_{i=1}^{C}\left [  \bar{\ell } \left ( \bar{y}_i ^s,\bar{y} _i ^w \right ) +  \bar{\ell } \left ( \bar{y}_i ^w,\bar{y} _i ^s \right ) \right ] -\eta (Y^\Lambda)  
\end{equation}
where the $sim(u,v)=u^Tv/\left \| u \right \| \left \| v \right \| $ denotes the cosine similarity for column vector. $\eta (Y^\Lambda) $ represents the regularization term designed to mitigate the risk of potential mode collapse during optimization, a phenomenon where all samples are inadvertently mapped to a single category \cite{r67}. This regularization is grounded in the concept of classification entropy, which quantifies the diversity of the classification outcomes:
 \begin{equation}
\eta (Y^\Lambda)=-\sum_{i=1}^{C} \left [ \left \| \bar{y} _i ^\Lambda  \right \|_1 log(\left \| \bar{y} _i ^\Lambda  \right \|_1)\right ]
\end{equation}

Overall, the schematic diagram of the aforementioned dual-contrast mechanism for the representation matrix of a mini-batch spectral data can be seen in Fig. 1(e). It optimizes the representation matrix using the intrinsic properties of spectral data from both row and column perspectives, making it more discriminative without the need for labels.

\subsubsection{Self Calibration}

For unannotated samples $x_i^U$, the category head can be employed to estimate their classification probabilities. When the model's predictions reach a high level of confidence, these predicted categories can be utilized as pseudo-labels for the samples \cite{r68}, thereby facilitating self-calibration of the model as shown in Fig. 1(f).

Given that each weakly augmented spectra  $x_i^U(w)$ only undergoes minor local perturbations, they are more likely to yield reliable predictive outcomes. Therefore, we first calculate the distribution of predicted class probabilities as $\tilde{y} _i^U (w)= \Gamma _{W3}\left ( \Theta_{W1}\left ( x_i^U(w) \right )  \right ) $. Subsequently, we utilize high-confidence predictions as pseudo-labels:
 \begin{equation}
 \hat{y} _i^U =\mathbbm{1}_{[\max(\tilde{y} _i^U (w)) \geq \epsilon ]}  \mathrm{argmax}(\tilde{y} _i^U (w)))
\end{equation}
where the $\epsilon $ is a hyper-parameter representing the confidence threshold for selecting pseudo-labels. Based on these pseudo-labels, we can perform self-calibration in the aforementioned embedding and category space.

Firstly, self-calibration is carried out in the embedding space. We transform the self-supervised contrastive loss in the embedding space into a supervised contrastive loss \cite{r69} employing pseudo-labels. In this scenario, samples with the same pseudo labels are considered positive pairs, while samples with different pseudo labels are regarded as negative pairs. Consequently, the enhanced Eq. (3) and Eq.(4) can be re-expressed as:
 \begin{equation}
\ell' \left (z_i^s\right )  = -\mathrm{log} \frac{\sum_{j=1}^{B}\mathbbm{1}_{[ \hat{y} _j^U= \hat{y} _i^U]}\mathrm{exp}\left ( \mathrm{sim} \left ( z_i^s,z_j^\Lambda \right ) /\tau_e \right )  }{\sum_{j=1}^{B}\mathbbm{1}_{[ \hat{y} _j^U\ne \hat{y} _i^U]} \mathrm{exp}\left ( \mathrm{sim} \left ( z_i^s,z_j^\Lambda \right ) /\tau_e \right )  }   
\end{equation}
 \begin{equation}
\mathcal{L}_{emb}' = \frac{1}{B} \sum_{i=1}^{B}\left [ \ell' \left (z_i^s\right )+  \ell' \left (z_i^w \right )\right ] 
\end{equation}

Similarly, we perform self-calibration in the category space by leveraging pseudo-labels to supervise the predictive behaviors for the strongly augmented views of the samples.
 \begin{equation}
\mathcal{L}_{pse}=\frac{1}{M} \sum_{i=1}^{M}\mathbf{H}\left (  \hat{y} _i^U\mid \tilde{y} _i^U (s)\right )
\end{equation}

\begin{algorithm}[t]
\caption{Training of SCDC}
 \SetKwInOut{Input}{input}
\SetKwInOut{KwOut}{Models}
\KwIn{Training  dataset $\mathbf{D}_{train}^L$,  $\mathbf{D} _{train}^U$}
 \KwOut{Encoder $\Theta_{W1}\left ( \cdot  \right )$, embedding head $\Phi_{W2}\left ( \cdot  \right ) $, classification head $\Gamma _{W3}\left ( \cdot  \right )$}

Learnable parameters  $\boldsymbol{\Omega} =\left \{ W1, W2,W3 \right \} $\;
Training Epoch  $\mathcal{E}$, Learning rate $\mu $\;

\For{$e=1,..., \mathcal{E}$}
{

\textbf{(a)}. \emph{Supervised training}\;
Randomly extract data batches from $\mathbf{D}_{train}^L$\;
Calculate loss: $\mathcal{L}_{sup} \gets \mathrm{Eq.(2)}$\;

\textbf{(b)}. \emph{Unsupervised training}\;
Randomly extract data batches from $\mathbf{D}_{train}^U$\;
Calculate loss: $\mathcal{L}_{uns} \gets \mathrm{Eq.(12)}$\;

Backpropagation to update $\boldsymbol{\Omega}$:
$ \boldsymbol{\Omega}^{e+1}\gets  \boldsymbol{\Omega}^{e}-\mu \Big( \frac{\partial\mathcal{L}_{semi}}{\partial \boldsymbol{\Omega}^{e}}\Big)
$
}
Optimized parameters $\boldsymbol{\Omega}^\ast  =\left \{ W1^\ast, W2^\ast,W3^\ast \right \}$\;
return $\Theta_{W1}^\ast, \Phi_{W2}^\ast , \Gamma _{W3}^\ast$
\end{algorithm}
\subsection{Objective functions of hybrid supervision}

The proposed framework possesses a nature of hybrid supervision, allowing it to operate in a completely unsupervised manner without the need for annotations. The loss function in this context is as follows.
 \begin{equation}
\mathcal{L}_{uns}= \mathcal{L}_{cat}+\mathcal{L}_{emb}'+\mathcal{L}_{pse}
\end{equation}

In the semi-supervised scenario with a small amount of annotated data, the loss function can be expressed as: 

 \begin{equation}
\mathcal{L}_{semi}= \mathcal{L}_{sup}+\mathcal{L}_{uns}
\end{equation}

The training procedure of the proposed method can be found in a more concise form in Algorithm 1. After training, we can utilize the encoder $\Theta_{W1}\left ( \cdot  \right )$ and category head $\Gamma _{W3}\left ( \cdot  \right )$ to perform  predictions on spectral samples in the test set $\mathbf{D} _{test}$.

\section{Results}
\subsection{Materials and datasets}
In our study, we adopt three widely studied bacterial Raman datasets, namely the Bacteria ID dataset \cite{r1}, the Bacteria Strain dataset \cite{r12}, and the Marine Pathogenic Bacteria dataset \cite{r38}.

The Bacteria ID is a large-scale benchmark dataset containing Raman spectra from 30 different bacterial species and yeasts, representing the most common infections encountered in intensive care units worldwide. The dataset is divided into three subsets: the reference dataset, the fine-tuning dataset, and the test dataset. The reference dataset comprises 2000 spectra from each of the 30 species, totaling 6,000 spectra, while both the fine-tuning and test datasets contain 100 spectra per species. In the original experimental setup, the reference dataset is used for pre-training the model, the fine-tuning dataset for further model refinement, and the fine-tuned model is then evaluated on the test dataset. To promote efficient annotation, our experiments exclude the reference dataset, utilizing only the fine-tuning dataset for model training.

The bacterial strain dataset is also a large-scale bacterial Raman spectroscopy dataset. It encompasses nine strains across seven distinct species. For the spectral analysis of each cellular sample, five varied laser acquisition durations were implemented. In our study, we focused on the subset characterized by a 10-second acquisition time, comprising a total of 2,186 spectral samples. We proceeded to randomly designate 60$\%$ of these samples for the training set, with the remaining 40$\%$ allocated to the test set.

The Marine Pathogen Bacteria dataset is designed for the identification of pathogenic bacteria in seafood and environmental samples using Raman spectroscopy. It includes eight bacterial strains isolated from the intestines of squid, with each strain generating approximately 150 Raman spectral samples. In total, 1,138 samples were collected, with 910 samples allocated to the training set and 228 samples to the test set. 

\begin{table}
\caption{The network architecture used in SCDC}
\label{table}
\setlength{\tabcolsep}{3pt}
\centering
\begin{tabular}{p{88mm}}
${\includegraphics[width=88mm]{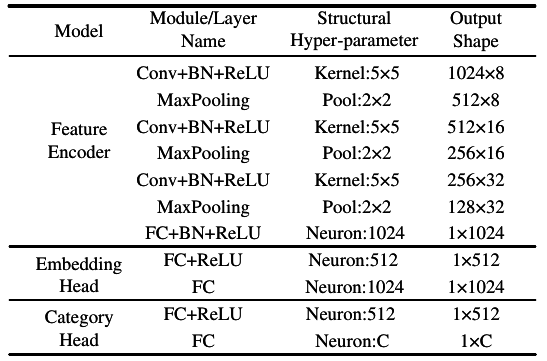}}$
\end{tabular}
\label{table_button}
\end{table}

\begin{table*}
\caption{The comparative results of the proposed SCDC method with existing advanced unsupervised \\clustering methods on three datasets.}
\label{table}
\setlength{\tabcolsep}{3pt}
\centering
\begin{tabular}{p{150mm}}
${\includegraphics[width=150mm]{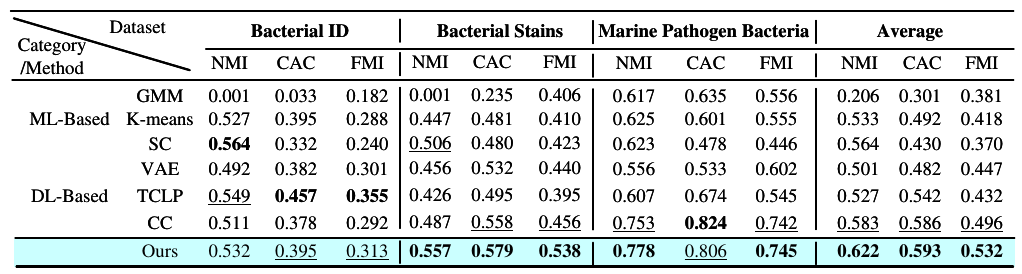}}$
\end{tabular}
\label{table_button}
\end{table*}

\subsection{Implementation details}

All experiments involving the proposed SCDC method and baseline approaches were implemented in PyTorch 1.12 and conducted on a server equipped with an NVIDIA 4090 GPU. During training, the Adam optimizer was employed with a learning rate of 1$e$-3 and a batch size of 32, without the use of weight decay or learning rate decay strategies. 
To ensure universality, the network architecture of SCDC follows a minimalistic design. Specifically, the encoder consists of a three-module convolutional neural network, with each module comprising a convolutional (Conv) layer, a batch normalization (BN) layer, a ReLU activation layer, and a max pooling layer. Both the embedding head and the classification head include an input fully connected (FC) layer, a ReLU activation layer, and an output fully connected layer. The detailed structural parameters are provided in Table 1.
Since different datasets collect spectral wavelengths over varying ranges, and the data lengths may differ, we interpolate the input spectra to a fixed length of 1024. Additionally, we apply max-min normalization to the spectra.

\subsection{Baselines}

In this study, several types of baseline models are considered, including unsupervised clustering models, semi-supervised classification models, and fully supervised classification models.

\subsubsection{Unsupervised clustering methods}
First, we consider unsupervised clustering methods, which do not require any annotations. We selected several representative methods for comparison, including traditional machine learning (ML) clustering methods such as Gaussian Mixture Model clustering (GMM), $k$-means clustering, and Spectral Clustering (SC), as well as deep learning (DL)-based representation clustering methods like Variational Autoencoder (VAE)\cite{r44}, Transfer Contrasting Learning Paradigm (TCLP)\cite{r65}, and Contrastive Clustering (CC)\cite{r70}. To ensure a fair comparison, the deep learning-based representation clustering methods use the same neural network structure as our approach. For VAE and TCLP, clustering results are obtained by applying $k$-means in the feature space extracted by the feature encoder.

\subsubsection{Semi-supervised classification methods}
Subsequently, we compared semi-supervised methods that only require partial annotation of the Raman spectroscopy data. We also selected representative methods from this category, including Pseudo-label\cite{r72} , Pi-Model\cite{r71}, MixMatch\cite{r73}, and FixMatch\cite{r74}. It is worth noting that these methods also do not involve any specific neural network design. For a fair comparison, we adopted the same neural network architecture as our method.

\subsubsection{Fully supervised classification methods}
The fully supervised model is also introduced as the fundamental intelligent model. We implemented basic deep neural network models, including a 4-layer CNN and a 4-layer LSTM model, as well as the state-of-the-art models ResNet\cite{r1} and SANet\cite{r16}, which are specifically designed for Raman spectral identification and have demonstrated significant performance.

\begin{table*}
\caption{The comparative results of the proposed SCDC method with existing advanced semi-supervised \\classification methods on the three datasets.}
\label{table}
\setlength{\tabcolsep}{3pt}
\centering
\begin{tabular}{p{150mm}}
${\includegraphics[width=150mm]{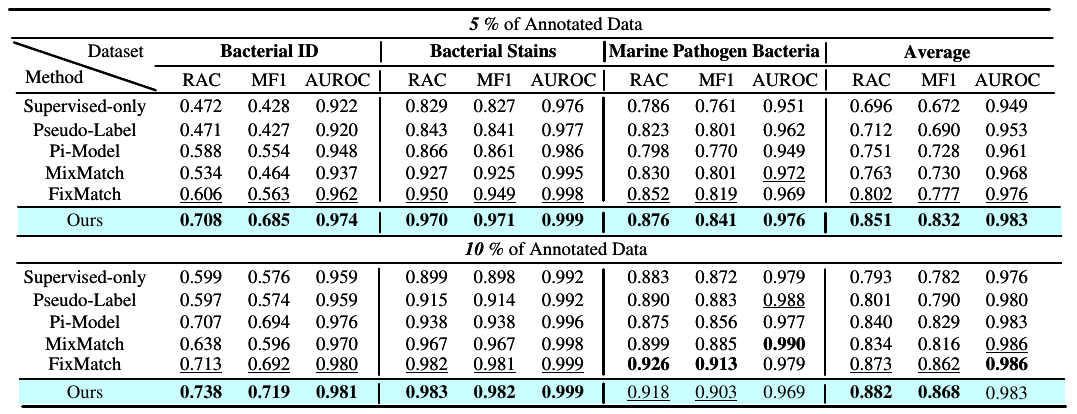}}$
\end{tabular}
\label{table_button}
\end{table*}

\subsection{Unsupervised results}
In this section, we conduct validation of unsupervised clustering, which does not require any data labels and is an important way to verify the model's ability to uncover the intrinsic properties of the data. In unsupervised clustering, we selected three commonly used metrics\cite{r70} to evaluate the clustering results, including Normalized Mutual Information (NMI), Clustering Accuracy (CAC), and Fowlkes-Mallows score (FMI).

Table II presents the comparative results of different methods on three bacterial Raman spectroscopy datasets, demonstrating that our method achieves robust performance across all three datasets. Out of the nine comparisons across the three datasets, our method attained the optimal or near-optimal results in eight instances. In contrast, other methods failed to achieve consistent performance across all three datasets. For example, the suboptimal CC method only achieved five optimal/near-optimal results, while TCLP achieved only three. In terms of average performance across the three datasets, our method ranked first in all three metrics. Compared to the second-best method CC in terms of average dataset performance, our method showed significant improvements, with gains of +3.9$\%$ in NMI, +0.7$\%$ in CAC, and +3.6$\%$ in FMI.

For ML-based approaches, we observed that GMM yielded the lowest performance, indicating that the data distribution of bacterial Raman spectroscopy is complex and cannot be characterized by simple Gaussian distributions, hence its poor performance. Although other traditional methods such as $k$-means and SC can effectively cluster bacterial Raman spectroscopy data, their performance is generally not robust.

For DL-based approaches, CC demonstrated overall significant results and performed well across the three datasets. TCLP, based on SimCLR, achieved good performance on the large-scale bacterial strain dataset, but its performance was average on the other two bacterial Raman spectroscopy datasets with smaller data volumes. VAE achieved acceptable results on the Marine Pathogen Bacteria dataset but exhibited degraded performance on the other two datasets, particularly on the complex Bacterial ID dataset.

\begin{table}
\caption{
The comparison of recognition accuracy between  fully supervised models using all annotations and our SCDC model using only a small amount of annotations.}
\label{table}
\setlength{\tabcolsep}{3pt}
\centering
\begin{tabular}{p{88mm}}
${\includegraphics[width=88mm]{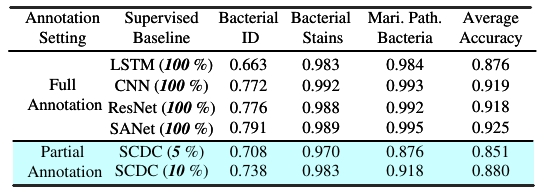}}$
\end{tabular}
\label{table_button}
\end{table}

\subsection{Semi-supervised results}

In the semi-supervised scenario, we only annotate a small number of samples, which can significantly reduce the labor of human experts and is the main scenario explored in this study. We investigated two annotation proportions: 5$\%$ and 10$\%$. We used recognition accuracy (RAC), F1 score (F1S), and Area Under the Receiver Operating Characteristic Curve (AUROC) to measure the identification accuracy of bacterial Raman spectroscopy. Both the F1S and AUROC were calculated using macro averaging.

Table III presents the performance of the proposed SCDC on three bacterial Raman spectroscopy datasets relative to five existing state-of-the-art semi-supervised methods. In terms of overall performance, our SCDC outperformed other methods in the average performance across the three datasets. Compared to the second-best method, FixMatch, our approach achieved improvements of +4.9$\%$/+5.5$\%$/+1.3$\%$ and +0.9$\%$/+0.6$\%$/-0.3$\%$ for RAC, F1S, and AUROC when using 5$\%$ and 10$\%$ of annotated data, respectively. Regarding the three metrics across the three datasets (equivalent to nine comparisons), our SCDC ranked first in all these metrics with 5$\%$ annotations and achieved eight optimal or near-optimal results with 10$\%$ annotations.

 \begin{figure}[t]
\centerline{\includegraphics[width=88mm]{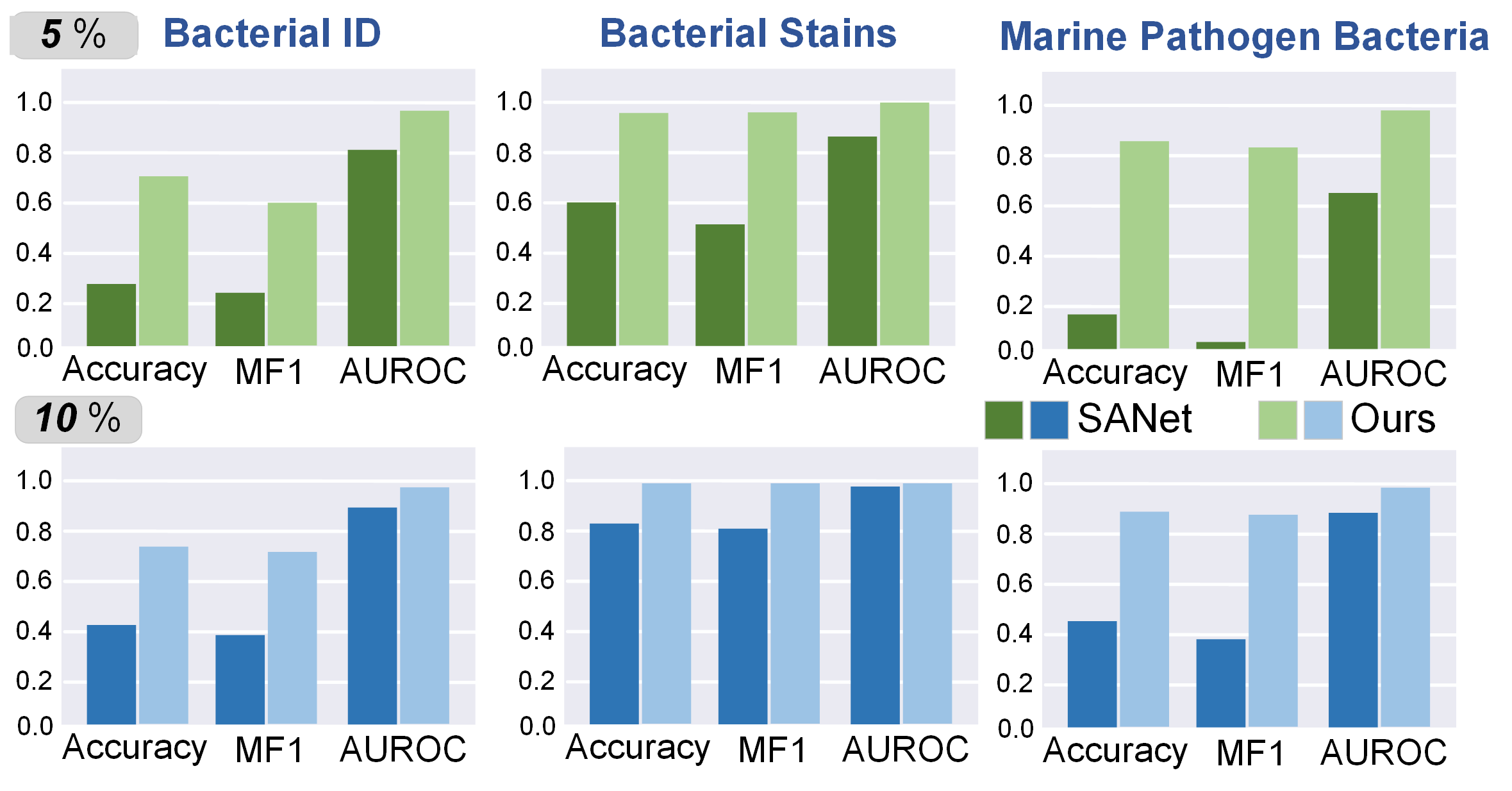}}
\caption[width=88mm]{Comparison of recognition accuracy between the advanced fully supervised model SANet and our proposed SCDC model under small proportion of annotation settings.

}
\label{fig1}
\end{figure}

From the perspective of annotation proportion, the performance gains of our method under the condition of 5$\%$ labeling are particularly significant, indicating that our approach can more effectively learn spectral features with a smaller amount of annotated data. When analyzing from the dataset perspective, it is evident that our method shows the most pronounced advantage on the most complex Bacterial ID dataset, highlighting its superiority in handling complex spectral recognition tasks.
\begin{table*}
\caption{
The ablation study results for different modules. the baseline model is trained using only supervised loss, and subsequent modules are added on this basis. The numbers in parentheses represent the improvement relative to the baseline.
}
\label{table}
\setlength{\tabcolsep}{3pt}
\centering
\begin{tabular}{p{180mm}}
${\includegraphics[width=180mm]{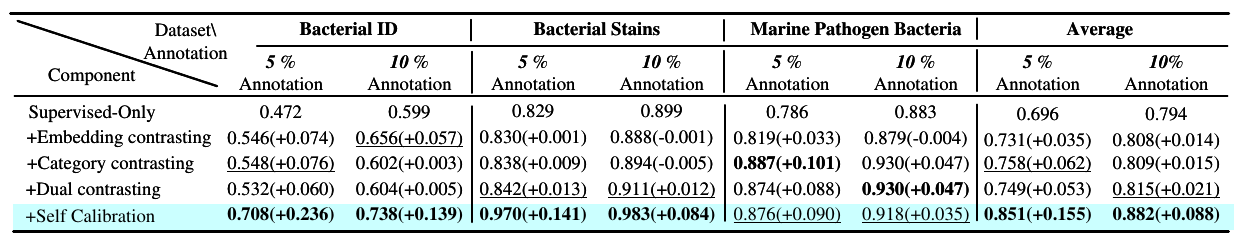}}$
\end{tabular}
\label{table_button}
\end{table*}

Furthermore, we compared the SCDC method with fully supervised methods that use all annotated data, and the results are presented in Table IV. It can be observed that our method achieves performance comparable to fully supervised methods on the Bacterial ID and Bacterial Strains datasets even when using only a small proportion(5$\%$ or 10$\%$) of annotated data, falling short only on the Marine Pathogen Bacteria dataset.

However, these fully supervised methods rely on a large amount of expert annotation. We compared the performance of the state-of-the-art fully supervised model, SANet, under the same low labeling proportion with our method, and the results are shown in Fig. 2. It can be observed that a small number of annotations can lead to over-fitting in such models, preventing them from obtaining a robust model with generalization capabilities, thus resulting in a significant degradation in performance. For instance, with only 5$\%$ annotations, SANet's recognition accuracy on the Bacterial ID and Marine Pathogen Bacteria datasets does not exceed 30$\%$, whereas, in contrast, our method achieves performance of 70.8$\%$ and 87.6$\%$, respectively.

The consistent superior performance across different datasets and varying proportions of annotated data demonstrates the effectiveness of SCDC in maximizing the utility of annotated data in a semi-supervised setting.

\begin{figure}[t]
\centerline{\includegraphics[width=88mm]{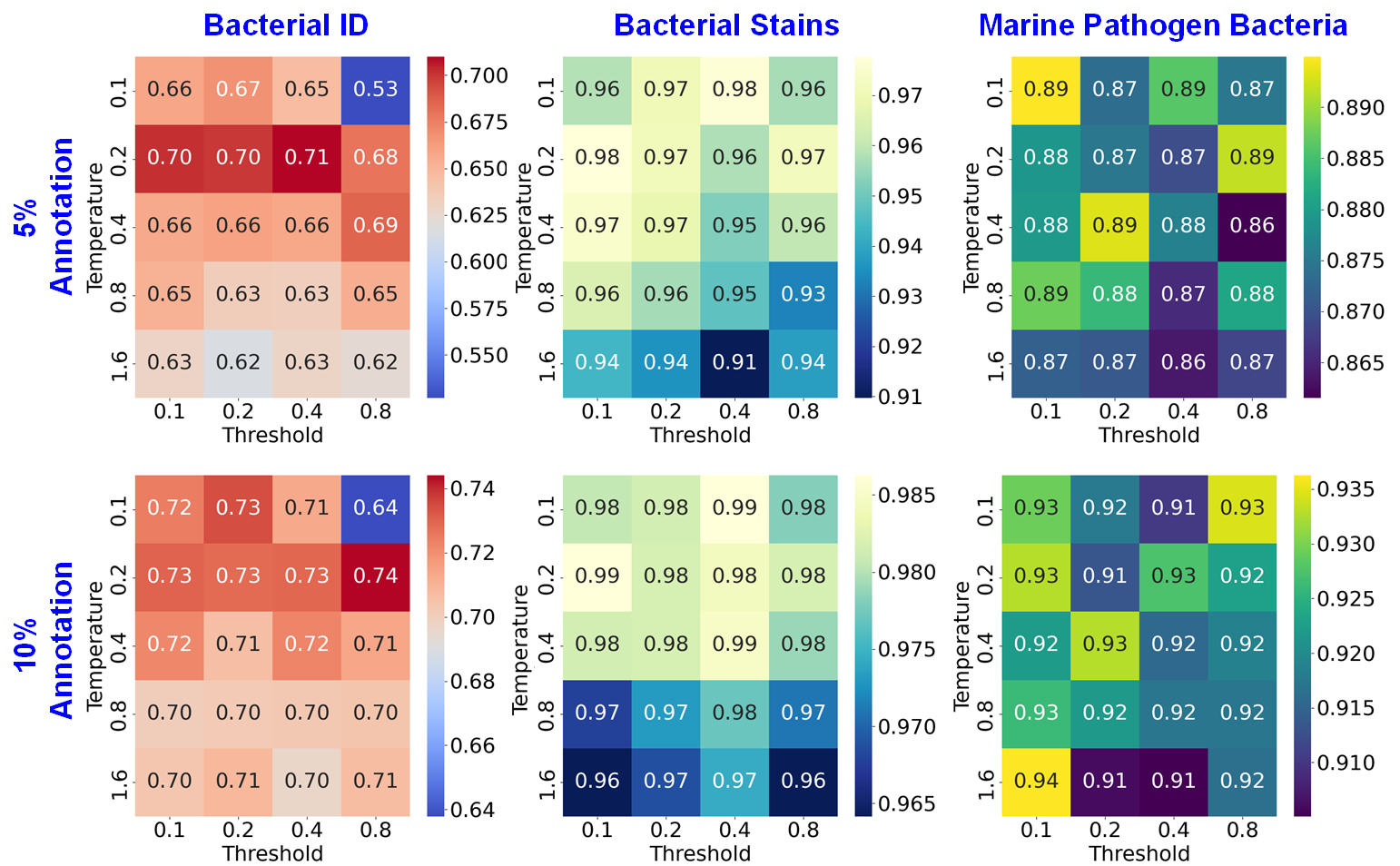}}
\caption[width=88mm]{
The heatmaps of recognition accuracy for temperature coefficient-threshold across three datasets.}
\label{fig1}
\end{figure}

\subsection{Analysis}

In this section, we will delve into the deep analysis of SCDC under limited annotation conditions to reveal its underlying mechanisms.

\subsubsection{Impact of dual contrasting}

Firstly, we analyze the impact of the dual-contrasting mechanism on recognition performance. This mechanism operates on the representation matrix of batch spectral data by analyzing row vectors and column vectors from two perspectives in the embedding and category subspaces, thereby enhancing the discriminability of features.

To investigate the impact of this mechanism on spectral classification, we conducted ablation experiments, the results of which are shown in Table V. We used a model trained with only supervised loss as the baseline, and the results indicate that incorporating the contrasting mechanism led to performance improvements across all three datasets, especially under the $5 \%$ annotation condition. When considering the effects of two contrasting mechanisms, it was observed that the improvement due to category contrasting was greater than that of embedding contrasting, and the dual contrasting mechanism outperformed category contrasting alone. On the whole, the introduction of the dual-contrast mechanism resulted in an average recognition accuracy improvement of $+5.3 \%$ and $+2.1 \%$ across the three datasets.

\begin{table}
\caption{The impact of different spectral augmentation \\ combinations on recognition accuracy}
\label{table}
\setlength{\tabcolsep}{3pt}
\centering
\begin{tabular}{p{88mm}}
${\includegraphics[width=88mm]{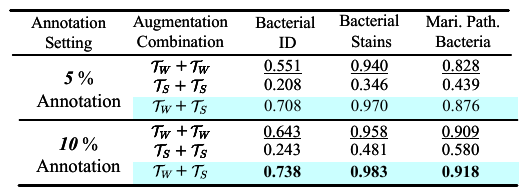}}$
\end{tabular}
\label{table_button}
\end{table}

\subsubsection{Gain from self-calibration}
Since the dual-contrast mechanism does not take into account any supervised information, the self-calibration was introduced to improve it, which employs the pseudo-supervision commonly used in semi-supervised learning. This mechanism is crucial for the model's performance. As shown in Table V, the dual contrasting enhanced with the self-correction mechanism achieved an average accuracy improvement of +10.2$\%$ and +6.7$\%$ across the three datasets. Furthermore, the dataset with a more complex distribution yields more substantial gains. For example, on the Bacterial ID dataset, the improvements were notably significant, with increases of +17.6$\%$ and +13.4$\%$.

 \begin{figure*}[t]
\centerline{\includegraphics[width=160mm]{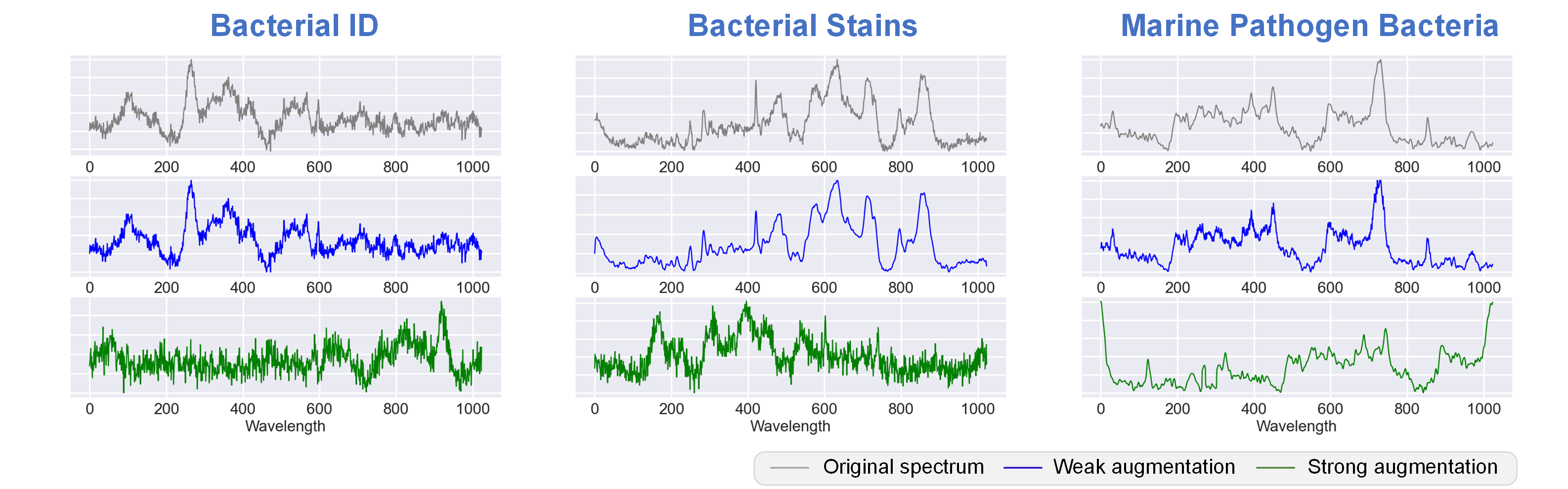}}
\caption[width=160mm]{
The spectral augmentation examples on the three datasets, where the first row shows the original samples, the second row displays the weak augmentation views, and the third row shows the strong augmentation views.
}
\label{fig1}
\end{figure*}

\begin{figure*}[t]
\centerline{\includegraphics[width=150mm]{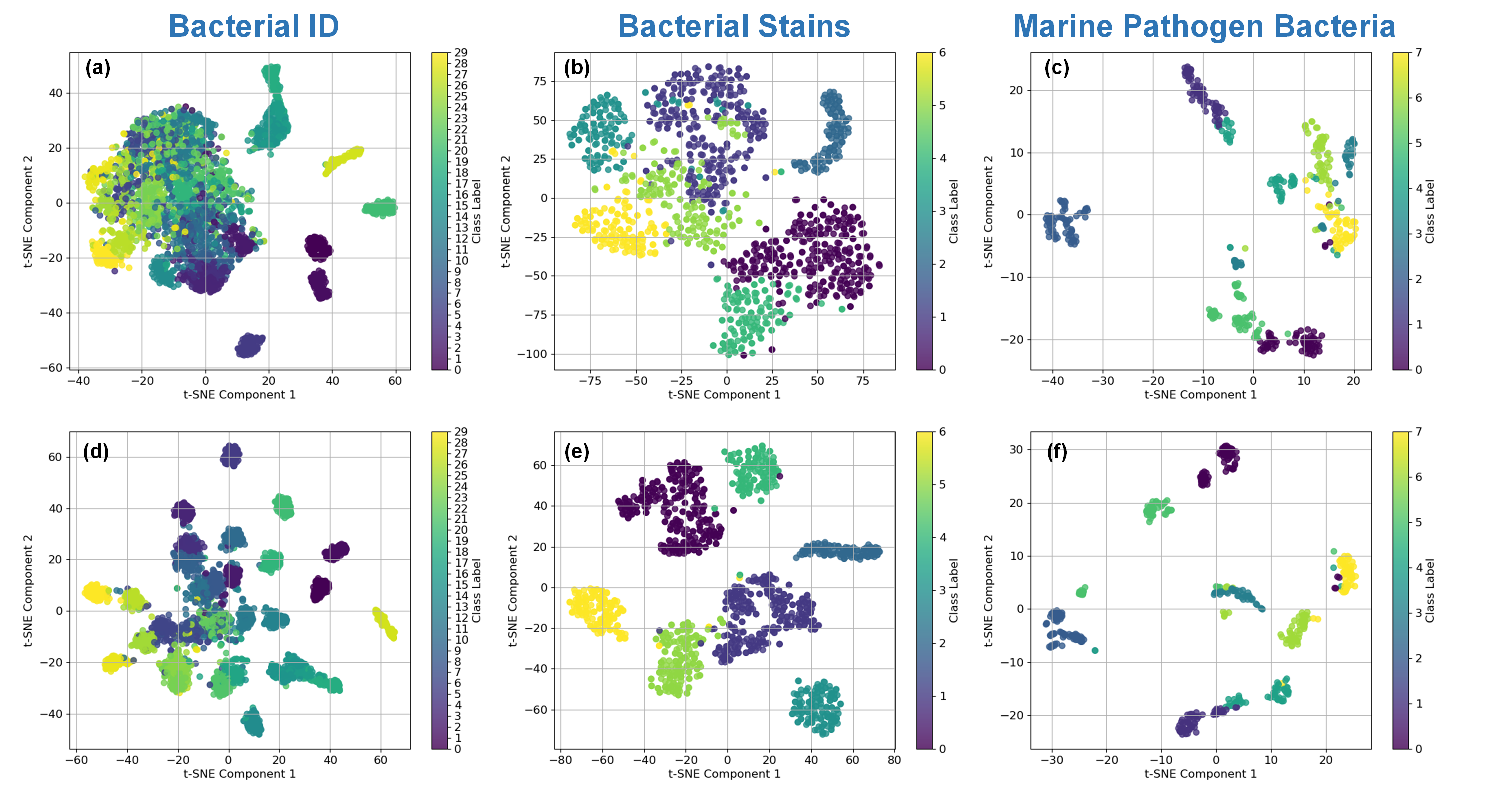}}
\caption[width=150mm]{
The learned feature representations' t-SNE visualization results on the three datasets. The first row represents the features learned by the baseline model, while the second row displays the features learned by the proposed SCDC. It is noteworthy that the visualization is performed on the test set.
}
\label{fig1}
\end{figure*}

\subsubsection{Trade-offs for Hyper-parameter}

In the proposed model, the two most critical hyper-parameters are the temperature coefficient $\tau$ and the threshold $\epsilon$. The temperature coefficient controls the smoothness of the feature distribution, while the threshold determines the reliability of pseudo-labels. Both have a significant impact on the extraction of discriminative features for spectral classification, which we have studied. The results of the threshold-temperature coefficient heatmap for classification accuracy are presented in Fig. 3. We investigated the temperature coefficient  $\tau$ ranging from 0.1 to 1.6 in multiple increments. The results indicated that a smaller $\tau$ generally yields better overall performance. However, a temperature that is too low can lead to instability in the model, as exemplified by the case of $\tau=0.1$, $\epsilon=0.8$ on the Bacterial ID dataset. Consequently, we set the default value of t to 0.2. The variation in the threshold $\epsilon$ exhibits less regularity, and the performance remains relatively stable with changes in $\epsilon$. Therefore, we selected 0.2 as the default value.

\subsubsection{Effect of spectral augmentation}
The selection of appropriate spectral augmentation methods is crucial for the proposed SCDC, as contrastive learning is highly sensitive to the construction of contrastive pairs\cite{r24}. In this study, we employed a strategy combining weak and strong augmentation. Here, we aim to investigate the impact of this strategy on Raman spectral identification.

This hybrid augmentation strategy takes into full consideration the physical significance of Raman spectroscopy, with representative examples across the three datasets shown in Fig. 4. Weak augmentation involves only local and limited amplification of the original spectral signal, such as the overall scale stretching, Gaussian smoothing, and random noise shown in the second row of Fig. 4. In contrast, strong augmentation has a semantic shift at the global level, such as the spectral shifting and flipping depicted in the third row of Fig. 4. The experimental results of different augmentation combinations are shown in Table VI. It can be observed that when using only weak augmentation, the model tends to focus on local semantic changes and fails to comprehend the global contextual semantics of the spectrum. When using only strong augmentation, the original semantic information of the spectrum is significantly disrupted, leading to a substantial decrease in performance. In contrast, the combination of both weak and strong augmentations allows the model to learn both local and global semantics adequately, resulting in the best performance.

\subsubsection{Visualizing learned representations}
Finally, we conducted the T-SNE\cite{r19} visualization of the feature representations learned by the model to intuitively display their discriminability, with the results shown in Fig. 5.

The first row of Fig. 5 displays the features learned by the baseline model using only supervised loss ($\mathcal{L}_{sup}$), while the second row shows the features learned by the model incorporated with  the self-calibrated dual contrasting mechanism ($\mathcal{L}_{sup}+\mathcal{L}_{uns}$). It can be observed that the discriminability of the features in the second row has been significantly enhanced, with the manifolds of different categories being more tightly clustered, whereas in the first row, features from different categories are intermingled. Consequently, the features obtained by our method exhibit stronger class discriminability, thereby enhancing the performance of spectral identification.

\section{Conclusion}

In this study, we address the issue of labor-intensive expert annotation in bacterial Raman spectroscopy identification by proposing a label-efficient Raman spectroscopy identification framework, SCDC, capable of operating in both unsupervised and semi-supervised settings. This framework incorporates a dual contrasting mechanism in embedding and category subspaces and employs pseudo-labeling for self-calibration. We have validated it on three large-scale bacterial Raman spectroscopy datasets, confirming its robust recognition performance. In the future, we will conduct further research on more robust self-supervised learning tasks for Raman spectroscopy to enhance the effective learning of its feature representations.

\bibliographystyle{ieeetr} 
\bibliography{reference}

\end{document}